\documentstyle[pra,aps,twocolumn,epsf]{revtex}

\begin{document}
\title{Two-photon imaging with thermal light}
\author{Alejandra Valencia, Giuliano Scarcelli, Milena D'Angelo and Yanhua Shih}
\address{Department of Physics, University of Maryland, Baltimore
County, \\ Baltimore, Maryland 21250} \maketitle\date{20 July,
2004}

\vspace*{-10mm}
\begin{abstract}
We report the first experimental demonstration of two-photon
imaging with a pseudo-thermal source. Similarly to the case of
entangled states, a two-photon Gaussian thin lens equation is
observed, indicating EPR type correlation in position. We
introduce the concepts of two-photon coherent and two-photon
incoherent imaging. The differences between the entangled and the
thermal cases are explained in terms of these concepts.
\end{abstract}
\maketitle

Two-photon imaging has been the subject of massive studies since
its first demonstration ten years ago \cite{belinsky,pittman}. In
a general sense two-photon imaging involves a joint detection of
two photons at distant space-time points. The radiation from a
source is split into two separate optical paths; an object
(aperture) is placed in one of the paths, but the spatial
information of the object is recuperated in a nonlocal fashion by
means of the second order correlation measurement. The effect of
two-photon imaging has been brought to the general attention by an
experiment that exploited the quantum entanglement nature of a
photon pair generated via Spontaneous Parametric Down Conversion
(SPDC) \cite{pittman}. In that experiment the signal and idler
radiation of SPDC were sent to two distant detectors. An aperture
and an imaging lens were placed in the signal arm of the optical
setup just before a bucket detector; there was no optical element
in the idler arm; however, by scanning the idler detector in the
plane defined by a two-photon Gaussian thin lens equation, a sharp
and magnified image of the aperture was obtained in the
coincidence counts, even though the single counting rates of both
detectors were fairly constant during the scanning. This
historical experiment was named ghost imaging due to the fact that
the image of the object is formed by photons that never actually
pass through the aperture. Further investigations on entangled
photons brought to the development of a new field, named
two-photon geometric optics \cite{pittman2}.

Recently it has been argued that classically correlated light
might mimic some features of the entangled photon pairs in
coincidence imaging setups \cite{bennink}. Notice that the
possibility of simulating the two-photon imaging features of
entangled states with classical sources was not ruled out by the
authors of the original ghost imaging experiment \cite{pittman}.
Both the theoretical work of Abourraddy et.al. \cite{abourraddy}
and the experimental investigation of Bennink et.al.
\cite{bennink} stimulated a very interesting debate about the role
of entanglement in two-photon coincidence imaging
\cite{abourraddy2,gatti,bennink2,rubin,dangelo,bjork}. In
particular Bennink et al. \cite{bennink} have experimentally
demonstrated the possibility of performing ``far-field"
coincidence imaging, i.e., measuring momentum correlation only,
with pairs of photons (pulses) classically correlated in momentum.
Very recently Gatti et.al. \cite{gatti2} have proposed thermal (or
pseudo-thermal) radiation as a classical source to perform
``near-field" coincidence imaging, which measures position
correlation (at least partially), in a specific optical setup.
Since then a great deal of attention on the subject has been
induced \cite{Zhu}.

In this paper we wish to present the first experimental
demonstration of two-photon ghost imaging with thermal radiation.
In particular, we show that a thermal source is able to simulate
one of the main features of entangled two-photon imaging: the
two-photon Gaussian thin lens equation, i.e., the EPR type
correlation in position is partially observable.  The expression
partially accounts for the reduced visibility of thermal
two-photon images (50\% constant background). Furthermore, we
introduce the concepts of two-photon incoherent and two-photon
coherent imaging to explain the fundamental differences between
thermal and entangled two-photon imaging. Notice that the thermal
ghost image presented here is the first authentic simulation of
the quantum ghost image \cite{pittman}. In fact, no two-photon
thin lens equation, which is a symbol of position correlation, can
be found by replacing SPDC with a source of photons classically
correlated in momentum \cite{bennink}. Thermal sources provide an
ideal comparison to quantum ghost imaging effects for a more
complete understanding of two-photon, or two-particle physics.

The behavior of entangled two-photon systems has been well studied
\cite{klyshko2,shih}. It is possible to establish an analogy
between classical optics and entangled two-photon optics: the
two-photon probability amplitude plays in entangled two-photon
processes the same role that the complex amplitude of the electric
field plays in classical optics; the role played by the intensity
of the electromagnetic field in classical optics is played by the
rate of coincidence counts, and therefore by the time integrated
second-order correlation function in entangled two-photon
processes. For both thermal source and SPDC, the two-photon thin
lens equation is the result of the coupling of two-photon
probability amplitudes. However, for an SPDC source, one pair of
photons contains all the two-photon probability amplitudes that
generate the ghost image; we define the result as
\textit{two-photon coherent} image. For a thermal source, instead,
the various two-photon probability amplitudes come from the
ensemble of many independent pairs that generates the ghost image;
thus, we define it as \textit{two-photon incoherent} image.

The experimental setup is shown in Fig.~\ref{setup}. After the
pseudo-thermal source \cite{martienssen,characterization}, a
non-polarizing beam splitter ($BS$) splits the radiation in two
distinct optical paths. In the reflected arm an object, with
transmission function $T(\vec{x_{1}})$, is placed at a distance
$d_{A}$ from the $BS$ and a bucket detector ($D_{1}$) is just
behind the object. In the transmitted arm an imaging lens, with
focal length $f$, is placed at a distance $d_{B}$ from the $BS$,
and a multimode optical fiber (then connected to detector $D_{2}$)
scans the transverse plane at a distance $d'_{B}$ from the lens.
The output pulses from the two single photon counting detectors
are then sent to an electronic coincidence circuit, to measure the
rate of coincidence counts.

The rate of coincidence counts is governed by the second order
Glauber correlation function \cite{Glauber}:
\begin{eqnarray}\label{G2}
& & G^{(2)}(t_{1},\vec{r}_{1}; t_{2},\vec{r}_{2})  \equiv \\
\nonumber \langle & &
E_{1}^{(-)}(t_{1},\vec{r}_{1})E_{2}^{(-)}(t_{2},\vec{r}_{2})
E_{2}^{(+)}(t_{2},\vec{r}_{2})E_{1}^{(+)}(t_{1},\vec{r}_{1})\rangle.
\end{eqnarray}
where $E^{(-)}$ and $E^{(+)}$ are the negative-frequency and the
positive-frequency field operators describing the detection events
at the space-time locations $(\vec{r_{1}}, t_{1})$ and
$(\vec{r_{2}}, t_{2})$. The transverse second order correlation
function (at equal times) for a thermal source is given by
\cite{next}:
\begin{eqnarray}\label{G2thermalg1g2}
G^{(2)}_{thermal}(\vec{x}_{1};\vec{x}_{2})&\propto&
\sum_{\vec{q}}|g_{1}(\vec{x}_{1},\vec{q})|^{2}\sum_{\vec{q}'}|g_{2}(\vec{x}_{2},\vec{q}')|^{2}
\nonumber\\&+&|\sum_{\vec{q}}g_{1}^{*}(\vec{x}_{1},\vec{q})g_{2}(\vec{x}_{2},\vec{q})|^{2},
\end{eqnarray}
where $\vec{x_{i}}$ is the transverse position of detector
$D_{i}$, $\vec{q}$ and $\vec{q'}$ are the transverse components of
the momentum vectors and $g_{i}(\vec{x}_{i};\vec{q})$ is the
Green's function associated with the propagation of the field from
the source to the $i^{th}$ detector \cite{rubincalcu}. Notice that
there are two differences with respect to the SPDC case: $(1)$ the
presence of a background noise (first term of
Eq.~\ref{G2thermalg1g2}), which does not exist for SPDC; $(2)$ the
possibility of writing the second term of Eq.~\ref{G2thermalg1g2}
as a product of first order correlation functions,
$G_{12}^{(1)}G_{21}^{(1)}$ while the second order correlation
function for SPDC, cannot be reduced to a mere product of first
order correlation functions in any format.

For the setup of Fig.~\ref{setup}, it can be shown that for any
values of the distances $d_{A}$, $d_{B}$, and $d'_{B}$ which obey
the equation:
\begin{eqnarray}\label{thinlens}
\frac{1}{d_{B}-d_{A}}+\frac{1}{d_{B}^{'}}=\frac{1}{f},
\end{eqnarray}
Eq.~\ref{G2thermalg1g2} can be simplified as \cite{next}:
\begin{eqnarray}\label{imagethermal}
G^{(2)}_{tot}(\vec{x}_{2}) \propto
N+|T(\frac{d_{A}-d_{B}}{d'_{B}}\vec{x}_{2})|^{2},
\end{eqnarray}
where $T(\frac{d_{A}-d_{B}}{d'_{B}}\vec{x}_{2})$ is the object
transmission function ($T(\vec{x_{1}})$) reproduced on the $D_2$
plane. We can then conclude that a thermal source allows
reproducing in coincidence measurements the ghost image of an
object, similarly to the SPDC case, except for a constant
background noise. The constant background noise, first term in
Eq.~\ref{imagethermal}, is proportional to the total transmittance
of the object and, therefore, can be simplified to N, the number
of transparent features in the object plane \cite{next}.
Eq.~\ref{thinlens} can clearly be interpreted as a two-photon
Gaussian thin lens equation, in which the object distance is
$s_{o}= d_{B}-d_{A}$ and the image distance is $s_{i}=d_{B}'$ (see
Fig.~\ref{unfold}).  We expect to observe an inverted image
magnified by a factor of $M=s_{i}/s_{o}$. Notice that when only
one slit is inserted in the object plane, the maximum achievable
visibility is $33\%$. The visibility, however, is expected to drop
when the number of transparent features increases.

The discovery of general laws in physics allows considering
pictorial representations which may turn out to be powerful but
still simple predictive tools. In this sense Klyshko's pictures
for SPDC imaging experiments \cite{klyshkobook,pittman} are
exemplar. The results presented in Eq.~\ref{thinlens} and
Eq.~\ref{imagethermal} also offer the possibility of considering a
generalized Klyshko's representation; nevertheless, the presence
of two terms in  Eq.~\ref{imagethermal} makes the thermal
Klyshko's picture more complex. For simplicity, in
Fig.~\ref{unfold} we only picture the second term of
Eq.~\ref{imagethermal}, which is the one that produces the ghost
image. Since the source is chaotic, each atom of the source
randomly emits photons with all possible values of momentum. A
photon with momentum $\vec{q}$ arriving at $\vec{x}_{1}$ can
produce a coincidence count with any other photon emitted by the
source and arriving at $\vec{x}_{2}$. These coincidences result in
a background noise that does not lead to the production of an
image. This part of the physical process is not reflected in
Fig.~\ref{unfold}. Notice that, similarly to the SPDC case, the
point-to-point correspondence between object and image plane is
the result of the addition of two-photon probability amplitudes.
It is important to point out however a remarkable difference
between the thermal and the SPDC case. In the Klyshko's picture
for SPDC each pair of lines corresponds to a two-photon
probability amplitude and all of them are associated to one pair,
hence we define the image \textit{ two-photon coherent}. In the
diagram for the thermal case each pair of lines represents the
term $g_{1}^{*}(\vec{x}_{1},\vec{q})g_{2}(\vec{x}_{2},\vec{q})$
and the various lines are associated to different pairs of
photons. It is apparent then that the thin lens equation is the
result of the coupling
$g_{1}^{*}(\vec{x}_{1},\vec{q})g_{2}(\vec{x}_{2},\vec{q})$, but
the image is generated by the ensemble of many independent pairs,
leading us to the definition of \textit{two-photon incoherent}
image. The diagram of Fig.~\ref{unfold} also shows that while in
the Klyshko's picture for SPDC the source behaves as an ordinary
mirror, here the thermal source behaves as a phase conjugated
mirror; this is due to the definition of the object distance
$s_o=d_B-d_A$. This interpretation indicates the presence of a
pseudo-object in the plane $\sigma$, as shown in
Fig.~\ref{unfold}; scanning this plane the pseudo-object was
actually observed (see also Fig.~\ref{pointbypoint}).

Our first experimental measurement was aimed to verify the
existence of a point-to-point correspondence between object and
image plane, as expected by the existence of a thin lens equation.
The setup is the same as that of Fig.~\ref{setup}, but we used the
$60 \mu m$-diameter input aperture of an optical fiber as the
object (whose output was then coupled to detector $D_{1}$). As a
preliminary measurement we studied the temporal second order
correlation function.  An optimized coincidence time window was
chosen accordingly.  For the actual spatial correlation
measurement, we collected three sets of data for three different
positions of the fiber in the object plane; in every measurement
we kept the position of the fiber fixed and scanned detector
$D_{2}$ in the transverse direction. The results are shown in
Fig.~\ref{pointbypoint}: any shift of the fiber in the object
plane causes a shift of the correlation function in the opposite
direction, in analogy to standard geometrical optics. In
particular Fig.~\ref{pointbypoint} shows that shifting the fiber
in the object plane by $2 mm$ the correlation function shifts by
$4.3 mm$. Hence, the magnification in the imaging plane is
$M_{meas}=2.15$, which is very close to the expected value
($M_{expect}=2.16$). The achieved visibility is $26\%$. The
results shown in Fig.~\ref{pointbypoint} clearly show the
point-to-point correspondence between object and imaging plane in
agreement with Eq.~\ref{thinlens} and Eq.~\ref{imagethermal}.

Then we placed a double slit in the object plane (center to center
separation $1mm$, slit width $0.2 mm$) and repeated the
measurement. The results are shown in Fig.~\ref{results}. As
expected, the single counts are flat, while the coincidence counts
reproduce the magnified image of the double slit. The visibility
drops to about $12\%$.

The experimental data show that the visibility of the two-photon
image drops with the number of transparent features in the object
plane, as predicted in Eq.~\ref{imagethermal}. This effect is
readily understood by inspecting Fig.~\ref{pointbypoint}: for each
feature in the object plane, the whole image plane shows besides
the expected image a non-negligible noise level. Hence, if in the
object plane there are simultaneously three features, in the image
plane we will observe the addition of the three graphs of
Fig.~\ref{pointbypoint}, as predicted by Eq.~\ref{imagethermal}.
This clearly indicates that the background noise increases with
the number of transparent features to image at the expenses of the
visibility.

The physics behind quantum ghost image and thermal ghost image can
be understood as follows: in both cases the image is the result of
the addition of two-photon probability amplitudes nevertheless
such probability amplitudes have very different origins in the two
cases.  In SPDC, a signal-idler pair is described by a pure state
given by the superposition of an infinite number of two-photon
probability amplitudes. On the other hand, thermal sources are
modelled as an incoherent statistical mixture of many pairs of
photons; the various two-photon probability amplitudes are
provided by the entire ensemble of photon pairs. The different
models come from the different processes involved in the
generation of SPDC and thermal radiation. SPDC comes from a
nonlinear coherent interaction and consequently it is impossible
to identify the birth place of a photon pair. This is why each
photon pair contains simultaneously all the possible transverse
momenta necessary for imaging. On the other hand, for a thermal
source each photon is generated from a random excitation of
independent atoms. This is in line with the interpretation of the
two Klyshko's pictures given above and clarifies what we mean by
\textit{two-photon coherent} and \textit{two-photon incoherent}
imaging.

In conclusion, we have presented the first experimental
demonstration of two-photon ghost imaging with thermal-like
sources. For the first time a two-photon Gaussian thin lens
equation is found for non-entangled sources, indicating the
existence of, at least partial, EPR type correlation in position.
We have also introduced the concepts of two-photon coherent and
two-photon incoherent imaging to describe the different physics
behind entangled and thermal ghost imaging. The already poor
visibility of the thermal ghost image (max $33\%$), drops rapidly
for complicated apertures; therefore, practical applications of
thermal ghost imaging rely on the development of a detection
scheme which is insensitive to the background noise. Such
detection scheme is now under development in our laboratory.

The authors would like to thank S. Thanvanthri, J. Wen and  M.H.
Rubin for everyday helpful discussions.  Y.H.S. specially thanks
S.Y. Zhu for stimulating discussions and encouragement. This
research was supported in part by NSF, ONR and NASA-CASPR program.

\begin{figure}[hbt]
\caption{Experimental setup. Source diameter $\sim 200 \mu m$; $a
= 125 mm$; $d_{A}=88 mm$; $d_{B}=212 mm$; $d'_{B}=268.5 mm$ $f=85
mm$; fiber tip diameter $=60 \mu$.} \label{setup}
\end{figure}

\begin{figure}[hbt]
\caption{Conceptual unfolded version of the optical setup. Object
plane and image plane obey a two-photon gaussian thin lens
equation and are defined by $s_{o}=d_{B}-d_{A}$ and
$s_{i}=d_{B}'$, respectively. In terms of Klyshko's picture, the
thermal source acts as a phase conjugated mirror forming a
pseudo-object in the $\sigma$ plane.} \label{unfold}
\end{figure}

\begin{figure}[hbt]
\caption{Normalized second order correlation function vs. position
of $D_{2}$. The measurement shows the point-to-point
correspondence between object and image planes. a) Tip of the
fiber in the object plane located in the position indicated by the
square. b) Tip of the fiber in the object plane located in the
position indicated by the circle (central position). c) Tip of the
fiber in the object plane located in the position indicated by the
triangle.} \label{pointbypoint}
\end{figure}

\begin{figure}[hbt]
\caption{Single and coincidence counts vs. transverse position of
$D_{2}$ in the image plane ($x_{2}$). Single counts of both
$D_{1}$ (hollow circles) and $D_{2}$ (full circles) are flat. The
coincidence counts (solid line with circles) show a magnified
image of the object.} \label{results}
\end{figure}

\centerline{\epsfxsize=2in \epsffile{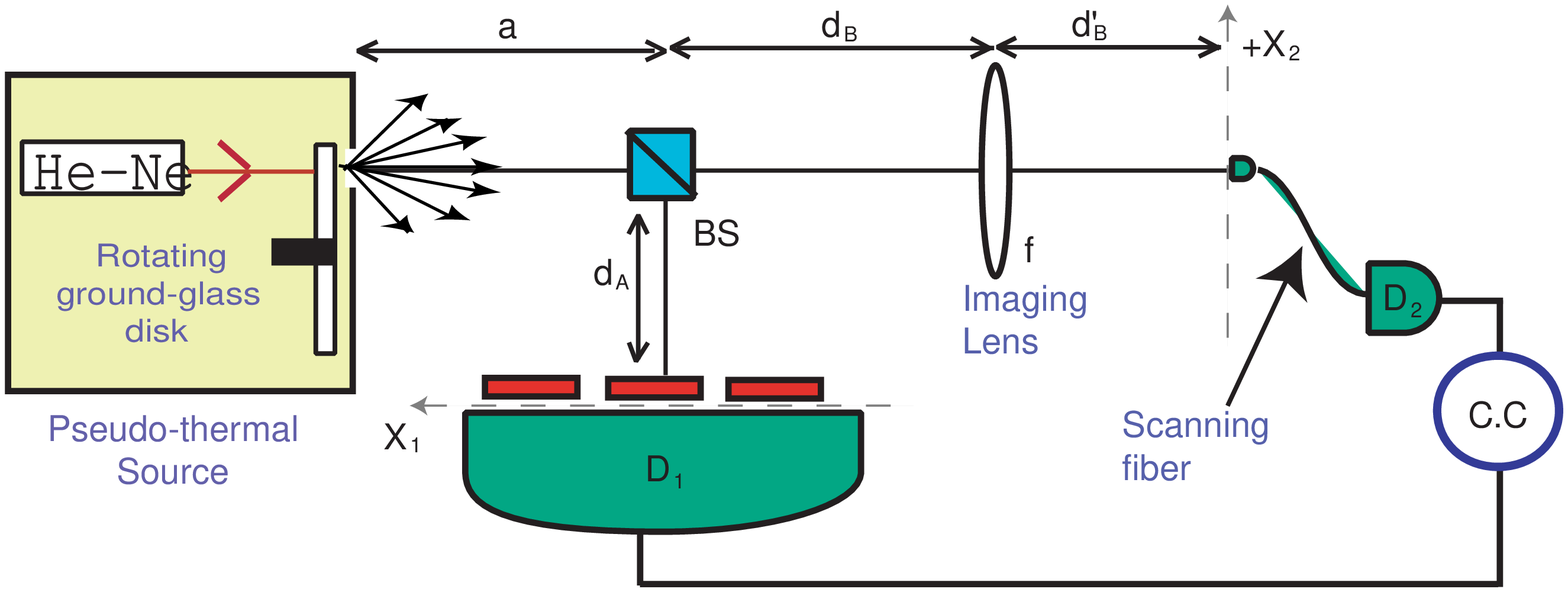}} Figure
\ref{setup}. Alejandra Valencia, Giuliano Scarcelli, Milena
D'Angelo, and Yanhua Shih.

\centerline{\epsfxsize=2in \epsffile{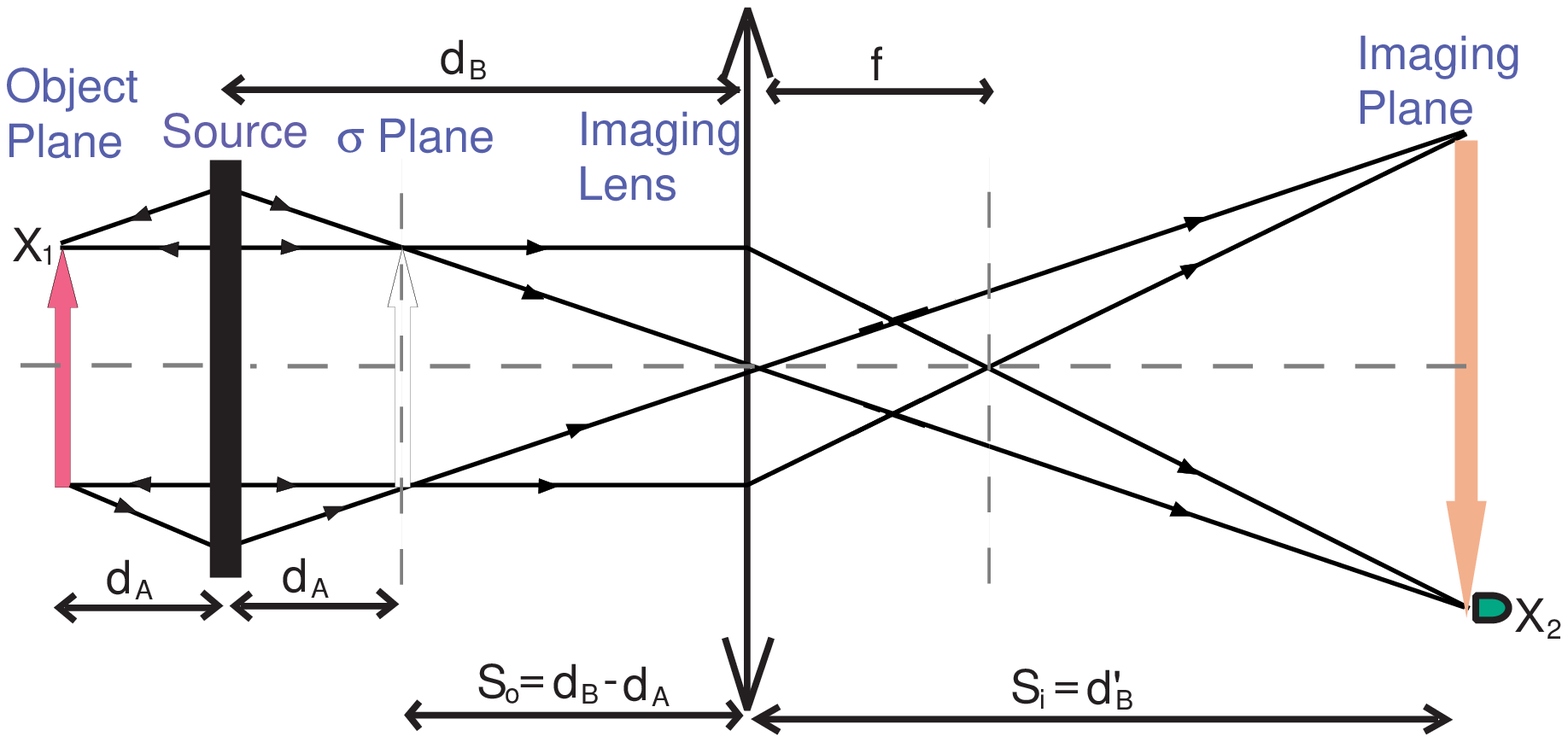}} Figure
\ref{unfold}. Alejandra Valencia, Giuliano Scarcelli, Milena
D'Angelo, and Yanhua Shih.

\centerline{\epsfxsize=2in \epsffile{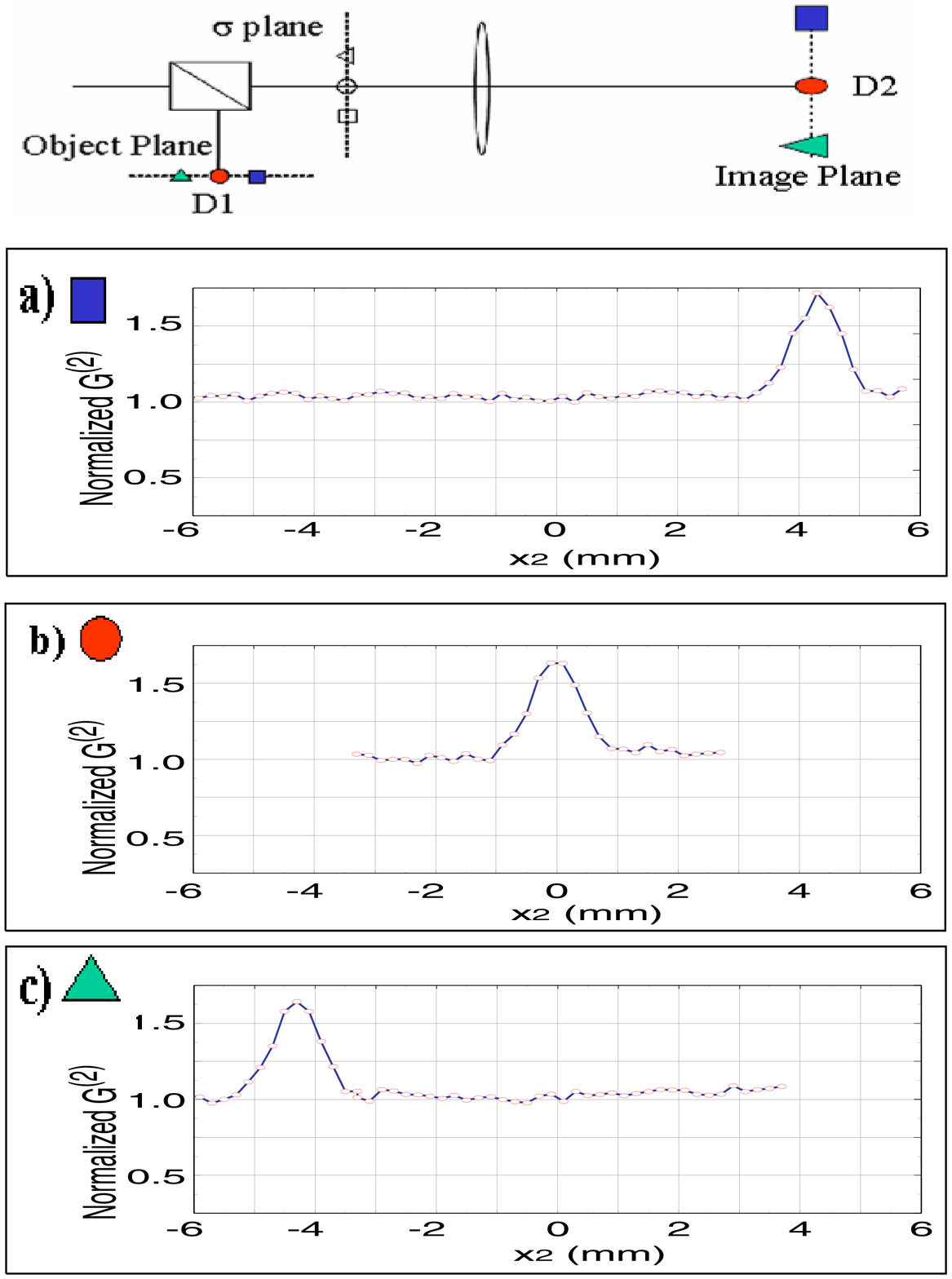}} \vspace{20mm}
Figure \ref{pointbypoint}.  Alejandra Valencia, Giuliano
Scarcelli, Milena D'Angelo, and Yanhua Shih.

\centerline{\epsfxsize=2in \epsffile{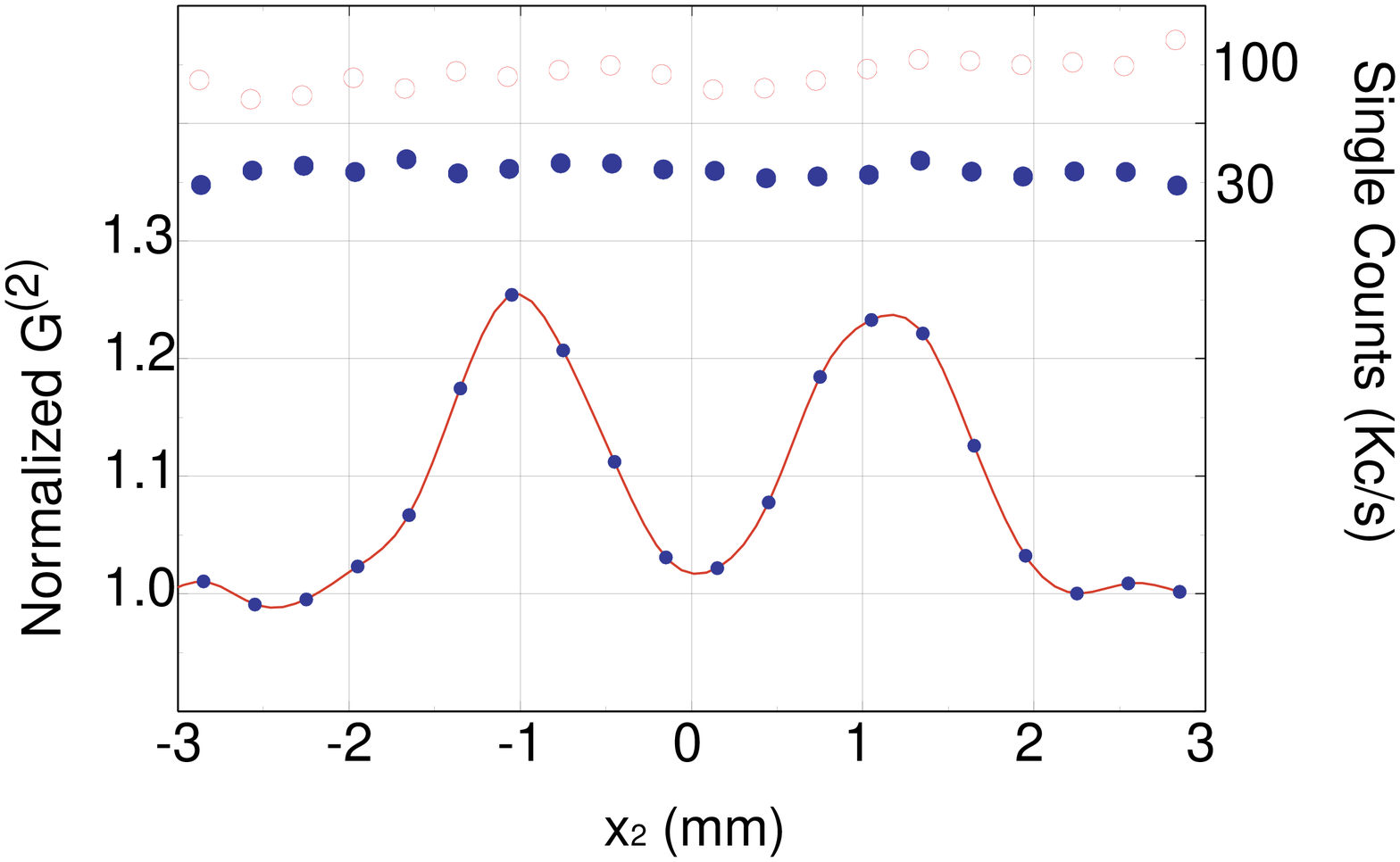}} Figure
\ref{results}.  Alejandra Valencia, Giuliano Scarcelli, Milena
D'Angelo, and Yanhua Shih.
\end{document}